%% file: v1-STAR-fxt-rev.tex
\documentclass[
prc,%
10pt,%
final,%
notitlepage,%
oneside,%
twocolumn,%
nobibnotes,%
nofootinbib,% In the current version of REVTeX, when this option is on,
superscriptaddress,%
floatfix]%
{revtex4}
\usepackage{color} %{\color{red} }
\usepackage{amsfonts} 
\usepackage{amsbsy} 
\usepackage{mathrsfs}
\usepackage{graphicx}
\usepackage{comment}

\def\lsim{\mathrel{\rlap{ \lower4pt\hbox{\hskip-3pt$\sim$}}
    \raise1pt\hbox{$<$}}} %less than approx. symbol
\def\gsim{\mathrel{\rlap{ \lower4pt\hbox{\hskip-3pt$\sim$}}
    \raise1pt\hbox{$>$}}} %greater than or approx. symbol
\def\scr#1{\mbox{\scriptsize #1}}

\begin{document}
% ============================================================

%\title{What Can We Learn from Directed Flow at STAR-FXT Energies?} 
\title{Examination of STAR fixed-target data on directed flow at $\sqrt{s_{NN}}=$ 3 and 4.5 GeV}

\author{Yu. B. Ivanov}\thanks{e-mail: yivanov@theor.jinr.ru}
\affiliation{Bogoliubov Laboratory of Theoretical Physics, JINR Dubna,
  141980 Dubna, Russia} 
\affiliation{National Research Center
  "Kurchatov Institute", 123182 Moscow, Russia}
\author{M. Kozhevnikova}\thanks{e-mail: kozhevnikova@jinr.ru}
\affiliation{Veksler and Baldin Laboratory of High Energy Physics,
  JINR Dubna, 141980 Dubna, Russia}

\begin{abstract}
We present results of simulations of directed flow of various hadrons  
in Au+Au collisions at collision energies of $\sqrt{s_{NN}}=$ 3 and 4.5 GeV. 
Simulations are performed within 
the model three-fluid dynamics and the event simulator based on it (THESEUS). 
The results are compared with recent STAR data. 
The directed flows of various particles provide information on dynamics in various parts 
and at various stages of the colliding system depending on the particle. 
However, the information on the equation of state is not always directly accessible 
because of strong influence of the afterburner stage or insufficient equilibration of the matter.  
It is found that the crossover scenario gives the best overall description of the data. 
This crossover equation of state is soft in the hadronic phase. 
The transition into QGP in Au+Au collisions occurs at collision energies between 3 and 4.5 GeV,  
at baryon densities $n_B \gsim 4 n_0$ and temperatures $\approx 150$ MeV.  
In-medium effects in the directed flow of (anti)kaons are discussed. 
%
%  \pacs{25.75.-q, 25.75.Nq, 24.10.Nz} 
%	\keywords{relativistic heavy-ion collisions, hydrodynamics, directed flow}
\end{abstract}
\maketitle
% \today

% ______________________________________________________________________
\section{Introduction}

The directed  flow is  
one of the most sensitive quantities to the dynamics of nucleus-nucleus collisions and properties of the matter  
produced in these collisions. It provides information about the stopping power of 
the nuclear matter, its equation of state (EoS), transition to quark-gluon phase (QGP) and more.  
All these issues were addressed in the analysis of the STAR data \cite{STAR:2014clz} obtained within 
Beam Energy Scan (BES) program at the Relativistic Heavy-Ion  Collider (RHIC). The analysis was performed 
within various approaches 
\cite{Konchakovski:2014gda,Ivanov:2014ioa,Ivanov:2016sqy,Steinheimer:2014pfa,Nara:2016phs,Shen:2020jwv,Ryu:2021lnx,Du:2022yok,Nara:2016hbg,Nara:2020ztb,Nara:2019qfd,Nara:2021fuu,Nara:2022kbb},
which include both hydrodynamic and kinetic models. An important conclusion of these studies is that 
the transition to the quark-gluon phase is most probably of the crossover or weak-first-order type 
and it stars at collision energies of $\sqrt{s_{NN}}<$ 8 GeV in Au+Au collisions. 
A promising recent development is prediction of correlation between the directed flow and 
the angular momentum accumulated in the participant region of colliding nuclei 
\cite{Ryu:2021lnx,Ivanov:2020wak,Tsegelnik:2022eoz,Jiang:2023fad,Jiang:2023vxp,Karpenko:2023bok}, which allows 
a deeper insight into collision dynamics.

The STAR-FXT (fixed-target) data on the directed flow of identified particles at energies 
$\sqrt{s_{NN}}=$ 3 and 4.5 GeV were recently published 
in Refs. \cite{STAR:2020dav,STAR:2021yiu}. 
%in Refs. \cite{STAR:2020dav,STAR:2021yiu,Liu:2023tqz}. 
These data were also analyzed within various, mostly kinetic models  
\cite{Nara:2020ztb,Nara:2022kbb,Oliinychenko:2022uvy,Steinheimer:2022gqb,OmanaKuttan:2022aml,Li:2022cfd,Wu:2023rui,Parfenov:2022brq,Mamaev:2023yhz,Yao:2023yda,Kozhevnikova:2023mnw,Kozhevnikova:2024itb,Yong:2023uct,Wei:2024oca} 
in relation to various problems: the hyperon production \cite{Nara:2022kbb,Wu:2023rui,Wei:2024oca}, 
the production of light (hyper)nuclei \cite{Kozhevnikova:2023mnw,Kozhevnikova:2024itb}, etc. 
The EoS of the matter produced in the nucleus-nucleus collisions was the prime topic of 
the above theoretical considerations. It was discussed mostly in terms of softness and 
stiffness of the EoS 
\cite{Nara:2020ztb,Oliinychenko:2022uvy,Steinheimer:2022gqb,OmanaKuttan:2022aml,Wu:2023rui,Yao:2023yda}. 
These studies were performed within different transport models: 
The relativistic version of the quantum molecular dynamics implemented into the
transport code JAM \cite{Nara:2020ztb},  the hadronic transport code SMASH \cite{Oliinychenko:2022uvy,Yao:2023yda}, 
the Ultrarelativistic Quantum Molecular Dynamics (UrQMD) \cite{Steinheimer:2022gqb,OmanaKuttan:2022aml}, and 
a multi-phase transport model \cite{Wu:2023rui}.

All the aforementioned papers 
\cite{Nara:2020ztb,Oliinychenko:2022uvy,Steinheimer:2022gqb,OmanaKuttan:2022aml,Wu:2023rui,Yao:2023yda}
reported that stiff (to a different extent) EoSs are preferable for the reproduction of the directed flow ($v_1$) 
at $\sqrt{s_{NN}}=$ 3 GeV, while the $v_1$ data at 4.5 GeV require a softer EoS. 
The latter was interpreted as indication of onset of the phase transition into QGP. 
This conclusion about preference of the stiff EoS at the energy of 3 GeV appears to 
contradict the earlier findings.
The analysis of KaoS \cite{KAOS:2000ekm} and FOPI \cite{FOPI:2011aa} data 
at collision energies $E_{lab} \leq 2A$ GeV ($\sqrt{s_{NN}} \leq$ 2.7 GeV) within the 
Isospin Quantum Molecular Dynamics model 
led to the conclusion that the soft EoS with the incompressibility $K =$ 210 MeV is strongly preferable 
\cite{FOPI:2011aa,Fuchs:2000kp,Hartnack:2005tr,Hartnack:2011cn}. 
Although, this energy range is somewhat below of the STAR-FXT one.

The energy range of the BNL Alternating Gradient Synchrontron (AGS), $E_{lab}$ = 2$A$--10.7$A$ GeV 
($\sqrt{s_{NN}} =$ 2.7-4.9 GeV), practically coincide with the currently explored STAR-FXT range.  
The results of the analysis of the AGS data \cite{E895:2000maf,E895:2000sor}
are more controversial. Strong preference of the soft EoS was reported in Refs. 
\cite{Ivanov:2014ioa,Ivanov:2016sqy,Pal:2000yc,Danielewicz:2002pu,Russkikh:2006ae}. 
In Refs. \cite{Ivanov:2014ioa,Ivanov:2016sqy}, the EoS additionally softens at  $\sqrt{s_{NN}} >$ 4 GeV
because of onset of the deconfinement transition. 
However, in Ref. \cite{Sahu:2002ku} it was found 
that the best description of the data on the transverse flow is provided 
by a rather stiff EoS at 2$A$ GeV (NL3) while at
higher bombarding energies (4$A$–-8$A$ GeV) a medium EoS  ($K =$ 300 MeV) leads to 
better agreement with the data, while  
the differences in the soft-EoS and stiff-EoS transverse flows
become of minor significance at 4$A$–-8$A$ GeV. 
In Ref. \cite{Isse:2005nk}, the proton flow was found to be also independent of stiffness of the EoS, 
however provided the momentum dependence in the nuclear mean fields is taken into account.

As recent studies  
\cite{Nara:2020ztb,Oliinychenko:2022uvy,Steinheimer:2022gqb,OmanaKuttan:2022aml,Li:2022cfd,Wu:2023rui,Yao:2023yda,Yong:2023uct}
of the STAR-FXT $v_1$ data deduced comparatively stiff EoSs at $\sqrt{s_{NN}}=$ 3 GeV,  
some of them predicted comparatively low baryon densities ($n_B$) for onset of the denfinement transition. 
This transition was associated with the softening of the EoS required for $v_1$ reproduction 
at the energy of 4.5 GeV. In terms of the normal nuclear density $n_0$, the deduced transition densities 
are 3-4$n_0$ \cite{Oliinychenko:2022uvy}, 4$n_0$ \cite{OmanaKuttan:2022aml}, 
2.5$n_0$  \cite{Li:2022cfd}, 3-5$n_0$ \cite{Wu:2023rui}, $n_B>$ 2-3$n_0$ \cite{Yao:2023yda}, and 5$n_0$
\cite{Yong:2023uct}. 
The model of three-fluid dynamics (3FD) \cite{Ivanov:2005yw,Ivanov:2013wha} predicts that 
the denfinement transition starts at approximately $n_B>$ 4-5$n_0$ at temperatures 100--150 MeV 
for the crossover EoS. 
However, the STAR-FXT data on the directed flow of identified particles at energies 
$\sqrt{s_{NN}}=$ 3 and 4.5 GeV have not yet been fully considered within the 3FD model, with the exception of
the proton and $\Lambda$-hyperon data at 3 GeV, which were analyzed with respect to the light 
(hyper)nuclei production \cite{Kozhevnikova:2023mnw,Kozhevnikova:2024itb}.

In view of the above reviewed developments, 
the debate about the EoS stiffness and onset of the QGP transition is far from being completed. 
A more extended discussion of the EoS constraints deduced from the directed-flow analysis 
can be found in recent review \cite{Sorensen:2023zkk}.

In the present paper, we present results of calculations of  the directed flow of various hadrons 
at energies $\sqrt{s_{NN}}=$ 3 and 4.5 GeV and compare them with recent STAR-FXT data 
\cite{STAR:2020dav,STAR:2021yiu}. The calculations are performed within the 3FD model 
\cite{Ivanov:2005yw,Ivanov:2013wha} and also within the
Three-fluid Hydrodynamics-based Event Simulator Extended by UrQMD final State interactions (THESEUS)
\cite{Batyuk:2016qmb,Batyuk:2017sku,Kozhevnikova:2020bdb}. 
The THESEUS simulations are intended to study the effect of the UrQMD afterburner stage on the directed flow. 
We present some conclusions that can be drawn from agreement or disagreement of the calculated results with the data.

%===================================================================
\section{3FD model and THESEUS generator} 
  \label{3FD and THESEUS}

The 3FD model \cite{Ivanov:2005yw,Ivanov:2013wha} simulates nonequilibrium at
the early stage of nuclear collisions by means of two counterstreaming
baryon-rich fluids. The third (fireball) fluid accumulates 
newly produced particles, dominantly populating the midrapidity region. 
These fluids, i.e. the projectile (p), target (t), and fireball (f), 
are governed by conventional hydrodynamic equations coupled by 
friction terms in the right-hand sides of the Euler equations. The friction terms 
describe the energy--momentum exchange between the fluids.

The hydrodynamic evolution ends with the freeze-out procedure described in Refs.
\cite{Russkikh:2006aa,Ivanov:2008zi}.  
The freeze-out criterion is $\varepsilon < \varepsilon_{\scr{frz}}$, 
where $\varepsilon$ is the total energy density of all three fluids in their common rest frame.
The freeze-out energy density $\varepsilon_{\scr{frz}}=0.4~$GeV/fm$^3$ 
was chosen mostly on the condition of the best reproduction 
of secondary particle yields for all considered EoSs, see \cite{Ivanov:2005yw}.  
The 3FD freeze-out includes an antibubble prescription, preventing formation of
bubbles of frozen-out matter inside the dense matter while it is still hydrodynamically evolving. 
The matter is allowed to be frozen out only if 
either (a) 
it is located near the border with the vacuum (this piece of matter gets locally frozen out) 
or (b) the criterion $\varepsilon < \varepsilon_{\scr{frz}}$
is met in the whole system (the whole system gets instantly frozen out).
The thermodynamic quantities of the frozen-out
matter are recalculated from the in-matter EoS, with which
the hydrodynamic calculation is performed, to the hadronic gas EoS.
This is done because a part of the energy is still accumulated in
collective mean fields at the freeze-out instant. 
This mean-field energy should be released before switching to the hadronic cascade 
in order to preserve energy conservation. 

The output of the model is recorded in terms of Lagrangian test particles 
(in terms of the numerical scheme ``particle-in-cell''), i.e. fluid droplets
for each fluid $\alpha$ (= p, t or f).
Each particle contains information on space-time coordinates %($t,\bf{x}$) 
of the frozen-out matter, proper volume of the test particle, 
%($V_\alpha^{\rm pr}$), 
hydrodynamic velocity, 
%($u^{\mu}_\alpha$) in the frame of computation, 
temperature, 
%($T_{\alpha}$), 
baryonic 
%($\mu_{\rm B\alpha}$)  
and strange 
%($\mu_{\rm S\alpha}$) 
chemical potentials. 
The THESEUS generator transforms the 3FD output 
into a set of observed particles, i.e. performs a particlization. 
%The particlization is followed by the afterburner stage. 

The 3FD model does not include any kinetic afterburner stage. 
The THESEUS event generator \cite{Batyuk:2016qmb,Batyuk:2017sku,Kozhevnikova:2020bdb}
does include the afterburner stage that is described by the UrQMD model.
The afterburner stage is of prime importance for   
collisions at lower energies, where there is no clear rapidity separation between
participant and spectator nucleons at the freeze-out. 
When the time
for the nuclei to pass each other becomes long relative to
the characteristic time scale for the participant evolution, the
interaction between participants and spectators (so-called
shadowing) becomes important \cite{Bass:1993ce,Heiselberg:1998es,Liu:1998yc}. 
In particular, 
the squeeze-out effect \cite{Sorge:1996pc,Danielewicz:1998vz,Ivanov:2014zqa}, 
is the consequence of this shadowing, i.e. results from blocking of the expanding 
central blob by the spectator matter.
This shadowing only partially is 
taken into account within the 3FD evolution because the central fireball 
remains to be shadowed even after the freeze-out while 
particles escape from this fireball without interacting with spectators in the 3FD model.

The afterburner stage should, in principle, correct this deficiency. 
However, it does not do it completely. 
The reason is that the THESEUS artificially assigns {\em the same time instant} to all produced particles
before proceeding to the afterburner, while different parts of the system are frozen-out at 
{\em different time instants} in 3FD. 
%The differential in time particlization would request the treatment of the interaction the frozen-out, kinetic 
%phase of the system with still hydrodynamically evolving matter. The latter is 
%a highly involved and ambiguous task. 
A time-extended transition from hydrodynamic evolution to afterburner dynamics would need treatment of 
the interaction of the kinetic afterburner phase with still hydrodynamically evolving matter. 
This is a difficult task both technically and conceptually.
The same time that is artificially assigned to all generated particles before 
the UrQMD stage is the way to avoid this difficulty, however on the expense of 
skipping this hydro-kinetic interaction.  
The lack of this interaction is the prime reason of shortcoming of the THESEUS afterburner.

At lower collision energies, participants are frozen out earlier than spectators.  
The spectators evolve slower because of the lower excitation energy and hence require longer time 
before the freeze-out. 
Therefore, the afterburner skips the stage of shadowing the afterburner expansion of the central fireball by spectators still being in the hydrodynamic phase.
It means that the evolution of the frozen-out participants is effectively stopped 
until the spectators also become frozen-out.
When the spectators also become frozen out,   
they have already partially passed the expanding central fireball.
Thus, the shadowing by spectators turns out to be reduced 
compared to what it would be if the entire collision process were 
kinetically treated, like in UrQMD or JAM.  

The 3FD model has been extensively used to simulations of Au+Au collisions at AGS energies, which 
almost coincide with the STAR-FXT ones. Quantities, which are low sensitive to the afterburner stage, 
were well reproduced by the 3FD simulations. These are various bulk observables 
\cite{Ivanov:2013wha,Ivanov:2013yqa,Ivanov:2013yla}, proton directed  
\cite{Ivanov:2014ioa,Ivanov:2016sqy} and  elliptic (at higher AGS energies) \cite{Ivanov:2014zqa} flow,   
bulk properties and directed flow 
of light (hyper)nuclei at $\sqrt{s_{NN}}=$ 3 GeV \cite{Kozhevnikova:2023mnw,Kozhevnikova:2024itb}. 
Problems with reproduction of the elliptic flow of protons 
and light nuclei at $\sqrt{s_{NN}}=$ 3 GeV in Ref. \cite{Kozhevnikova:2023mnw}
are related to the aforementioned deficiency of the isochronous particlization in THESEUS. 
Precisely the same parameters of the 3FD model as those in Refs.  
\cite{Ivanov:2014ioa,Ivanov:2016sqy,Kozhevnikova:2023mnw,Kozhevnikova:2024itb,Ivanov:2013wha,Ivanov:2013yqa,Ivanov:2013yla} are used in the present simulations.

\section{Equations of State} 
  \label{EoS}

\begin{figure}[!h]
%\vspace*{-14mm}
\includegraphics[width=5.6cm]{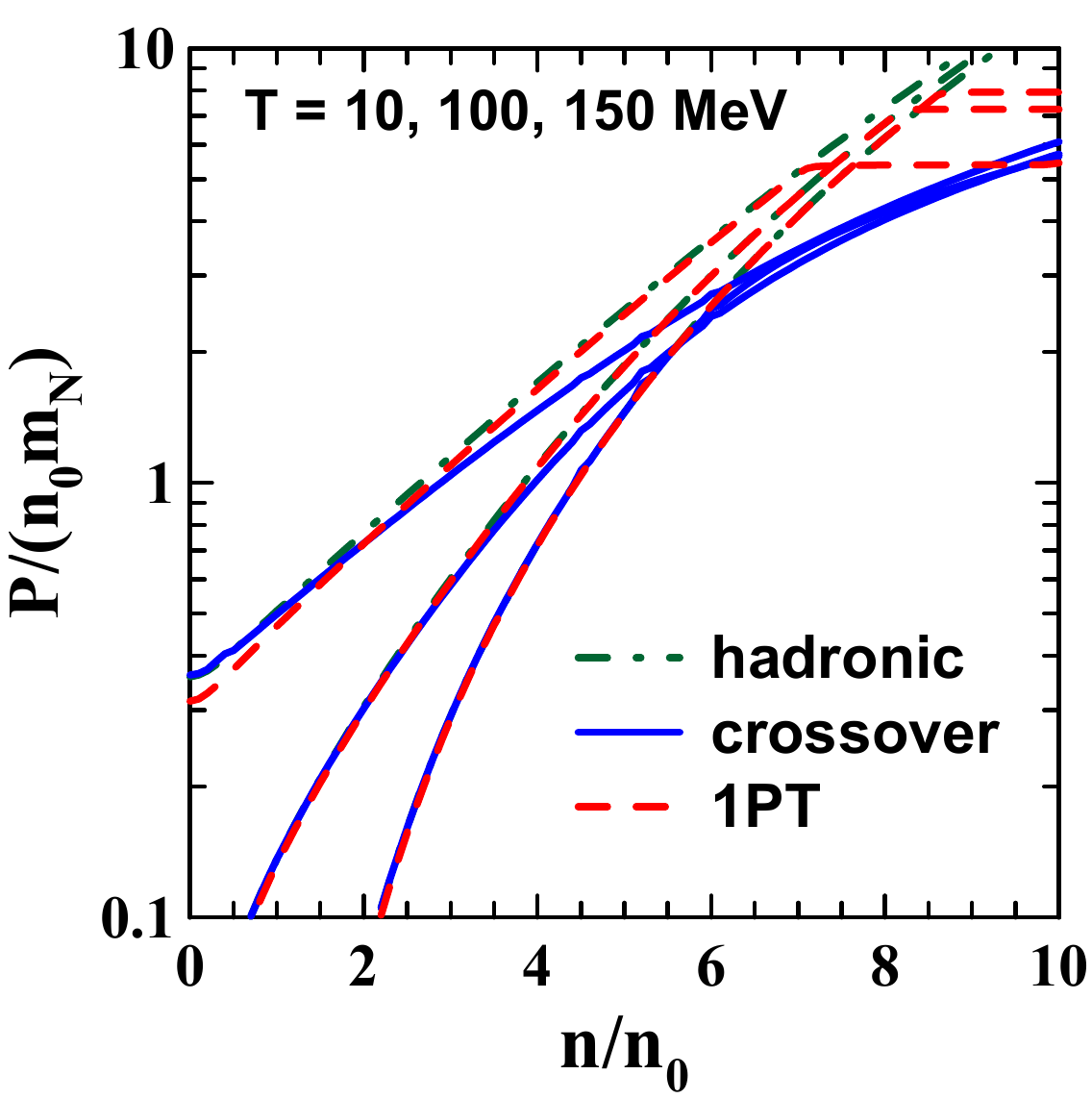}
 \caption{%(Color online)
Pressure (scaled by product of the normal nuclear density, $n_0=$ 0.15 1/fm$^3$, 
and the nucleon mass, $m_N$) at three temperatures, $T=$ 10, 100 and 150 MeV  
(from bottom upwards for corresponding curves), 
as function of the net baryon density (scaled by $n_0$) for the hadronic, crossover and 1PT EoSs.   
}
\label{fig:3-EOS-2024}
\end{figure}

\begin{figure}[!htb]
\includegraphics[width=6.2cm]{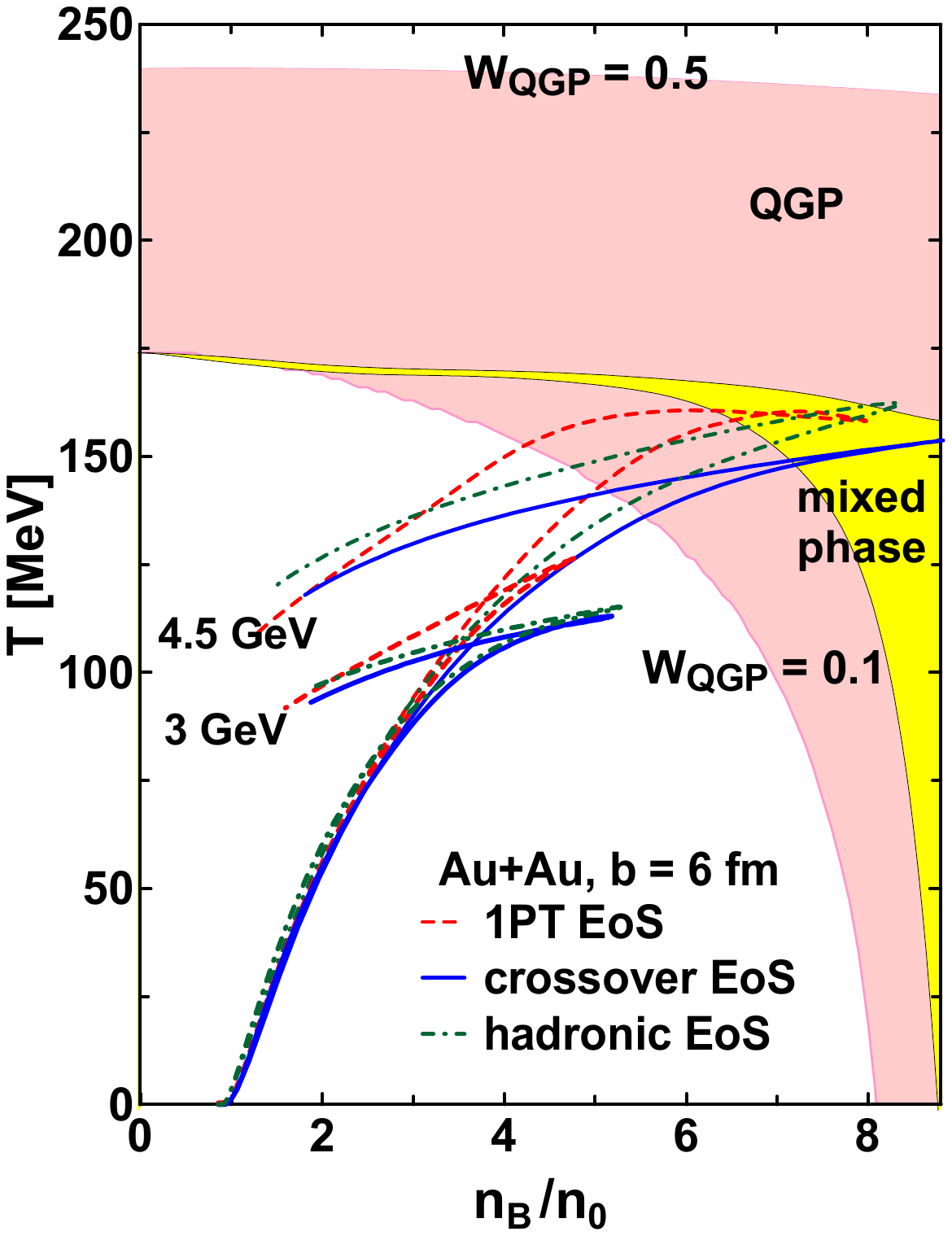}
 \caption{%(Color online)
Dynamical trajectories of the matter in the central cell of the colliding Au+Au nuclei  
in semicentral collisions (impact parameter is $b=$ 6 fm) at energies $\sqrt{s_{NN}}=$ 3  and 4.5 GeV. 
The trajectories are plotted in terms of the baryon density 
($n_B$, scaled by the normal nuclear density $n_0$) and temperature $T$. 
The trajectories are presented for the three EoSs. 
The mixed phase of the 1PT EoS is displayed by the shadowed region marked as ``mixed phase''. 
The wide shadowed area displays the region of the crossover EoS between the QGP fractions $W_{QGP}=$ 
0.1 and 0.5. 
}
\label{fig:TnB-box_2024}
\end{figure}

The 3FD model is designed to work with different EoSs. 
Three different EoSs are traditionally used in the 3FD simulations: 
a purely hadronic EoS \cite{gasEOS} and two EoSs
with deconfinement transitions \cite{Toneev06}, i.e. 
an EoS with a first-order phase transition (1PT EoS) and one with a
smooth crossover transition.  
While the hadronic EoS is quite flexible, i.e., it allows for changes of incompressibility, 
the  EoSs with deconfinement transitions are strictly tabulated. 
These EoSs are illustrated in Fig. \ref{fig:3-EOS-2024}.
As seen, all three EoSs are similar in the hadronic phase. Note that the displayed 
version of the hadronic EoS is characterized by incompressibility $K=$ 190 MeV. The simulations 
below are performed with this version of the hadronic EoS. The crossover pressure starts to deviate from 
the hadronic one at $n_B>$ 4-5$n_0$ at temperatures 100--150 MeV  
that are typical for the collisions at STAR-FXT energies, see Fig. \ref{fig:TnB-box_2024}.

Dynamical trajectories of the matter in the central cell of the colliding Au+Au nuclei  
in semicentral collisions ($b=$ 6 fm) at energies $\sqrt{s_{NN}}=$ 3  and 4.5 GeV 
are presented in Fig. \ref{fig:TnB-box_2024} in terms of the baryon density 
and temperature.
Evolution starts from the normal nuclear density and zero temperature and then follows an almost universal 
trajectory for some time. Shortly before reaching the turning point, at which density and temperature
are maximal, the matter in this central cell becomes equilibrated, as it was demonstrated in Ref. 
\cite{Ivanov:2019gxm}, and therefore the temperature takes its conventional meaning. 
The turning points at the energy of 3 GeV only touches the QGP region according to the 
crossover EoS, see Fig. \ref{fig:3-EOS-2024}. At the same time the 4.5-GeV trajectories fall well into  
the crossover QGP region and even enter the the 1PT mixed phase. 
The trajectories for different EoSs move away from each other at higher densities and temperatures. 
In particular, it demonstrates that the hadronic EoS and the 1PT one are not identical 
in the whole hadronic phase.

The crossover and 1PT phase diagrams require some comments. 
The QCD lattice calculations demonstrated that 
the transition into QGP at zero baryon chemical potential is a smooth crossover \cite{Aoki:2006we}. 
Due to that, the transition temperature is ambiguous because different definitions can lead 
to different values for it. Observables related to chiral symmetry result in the transition
temperature around 155--160 MeV \cite{Borsanyi:2020fev}. 
As seen from Fig. \ref{fig:TnB-box_2024},
the transition regions at zero baryon density (i.e. chemical potential) in EoSs of Ref. \cite{Toneev06} 
are located at noticeably higher temperatures than 155--160 MeV. 
This happens because the EoSs of Ref. \cite{Toneev06} were fitted to the old, 
still imperfect lattice data \cite{Fodor:2002sd,Csikor:2004ik,Karsch:2000kv}. 
Moreover, the crossover
transition constructed in Ref. \cite{Toneev06} is very smooth. The
hadronic fraction survives up to very high temperatures. In
particular, this is seen from Fig. \ref{fig:TnB-box_2024}: the fraction of the
quark-gluon phase ($W_{QGP}$) reaches value of 0.5 only at
very high temperatures. 
Such a smooth crossover is also used in the PHSD model (Parton-Hadron-String Dynamics) \cite{Cassing:2009vt}. 
However, this version of the crossover \cite{Toneev06} certainly contradicts results of the
lattice QCD calculations at zero chemical potential, where a fast crossover
was found \cite{Aoki:2006we}.
However, the aforementioned  shortcomings are not severe for 
the present simulations at relatively low collision energies, because the system evolution 
takes place in the region of high baryon densities, where the EoS is not known from the first principles.

\begin{figure}[!htb]
\includegraphics[width=7.6cm]{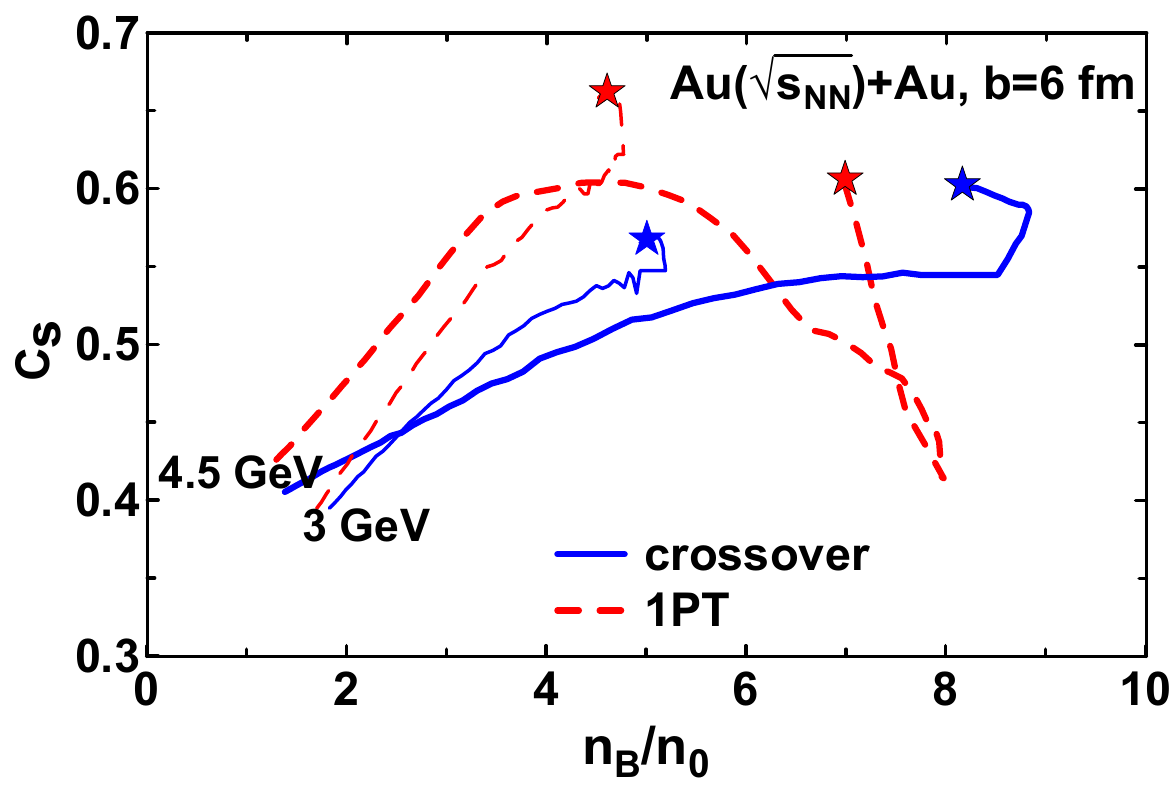}
 \caption{%(Color online)
Evolution of the isentropic speed of sound ($c_s$) as function of the 
baryon density ($n_B$, scaled by the normal nuclear density $n_0$) along 
the dynamical trajectories displayed in Fig. \ref{fig:TnB-box_2024}. 
The evolution is displayed from the instants (indicated by star symbols) 
when the matter is sufficiently equilibrated. 
The trajectories are presented for the 1PT EoS and crossover EoS. 
The trajectories for collisions at $\sqrt{s_{NN}}=$ 3 and 4.5 GeV are displayed by 
thin and thick lines, respectively. 
}
\label{fig:Cs-vs-nb-b6}
\end{figure}

In order to illustrate the difference between the 1PT EoS and crossover EoS in regions probed by 
the semicentral collisions at $\sqrt{s_{NN}}=$ 3 and 4.5 GeV, in Fig. \ref{fig:Cs-vs-nb-b6}  we present 
the evolution of the isentropic speed of sound ($c_s$) as function of the 
baryon density along the dynamical trajectories displayed in Fig. \ref{fig:TnB-box_2024}
\begin{eqnarray}
\label{Cs}
c_s = \left(\frac{\partial P}{\partial \varepsilon}\right)_{\rm{along \; trajectory}}, 
\end{eqnarray}
where $P$ and $\varepsilon$ are the pressure and energy density, respectively. 
We do not show $c_s$ for hadronic EoS to avoid overcrowding of the figure. 
The evolution is displayed beginning from instants (indicated by star symbols) 
when the matter is sufficiently equilibrated, i.e. the difference between the 
longitudinal  ($P_{\rm{long}}$) and transverse ($P_{\rm{tr}}$) pressures
\begin{eqnarray}
\label{P_long}
P_{\rm{long}}&=&T_{zz}, \hspace*{5mm}\mbox{(along the beam direction)}, 
\\
P_{\rm{tr}}&=&(T_{xx}+T_{yy})/2
\label{P_tr}
\end{eqnarray}
does not exceed 10\% \cite{Ivanov:2019gxm}. 
These pressures are defined in terms of the total energy--momentum tensor 
\begin{eqnarray}
\label{T_tot}
%T^{\mu\nu}_{\scr{tot}} \equiv
T^{\mu\nu} \equiv
T^{\mu\nu}_{\scr p} + T^{\mu\nu}_{\scr t} + T^{\mu\nu}_{\scr f}
\end{eqnarray}
being the sum of conventional hydrodynamical energy--momentum tensors of separate fluids
\begin{eqnarray}
\label{T_alpha}
%T^{\mu\nu}_{\scr{tot}} \equiv
T_\alpha^{\mu\nu} = (\varepsilon_\alpha + P_\alpha)
u_\alpha^\mu u_\alpha^\nu + g^{\mu\nu} P_\alpha, 
\end{eqnarray}
where $u_\alpha^\mu$ stands for the $\mu$-component of the 
hydrodynamic 4-velocity of the $\alpha$-fluid. 
The equilibration in the central region is attained 
shortly before reaching the turning point \cite{Ivanov:2019gxm}, at which density and temperature
are maximal, see Fig. \ref{fig:TnB-box_2024}. After that the evolution of the unified fluid is 
approximately (up to viscous-like dissipation) isentropic \cite{Ivanov:2016hes} and therefore Eq. (\ref{Cs})
takes the meaning of the isentropic speed of sound.

As seen from Fig. \ref{fig:Cs-vs-nb-b6}, the 1PT EoS and crossover EoS at the expansion stages of 
the semicentral collisions 
%at $\sqrt{s_{NN}}=$ 3 and 4.5 GeV 
are indeed different, which was indicated 
already in Fig. \ref{fig:TnB-box_2024}. It is remarkable that the softest-point region is probed at 
collisions at 4.5 GeV within the 1PT scenario. Indeed, the sound speed reaches minimum in the turning 
point of the 1PT trajectory. However, this softest-point region does not greatly affect the directed flow, 
as we will see below, since only the central region of the entire system falls within this softest-point region 
and only for a short time. The strong effect on $v_1$ occurs at higher collision energies, i.e. around 
energy of $\approx$8 GeV \cite{Ivanov:2014ioa,Ivanov:2016sqy}, when a large part of the matter  
falls within this softest-point region and for a longer time. 

At the same time, the softest point affects the midrapidity region of the 
rapidity distribution of net protons in central collisions  approximately at the same energy (4.9 GeV)
\cite{Ivanov:2013wha,Ivanov:2012bh,Ivanov:2015vna}. 
It happens because a more 
extended region falls within this softest-point range in central collisions and because the 
effect is located in the midrapidity that is closely related to the central region of the system.

\section{Directed Flow} 
  \label{Directed Flow}

\begin{figure*}[!tbh]
\includegraphics[width=.99\textwidth]{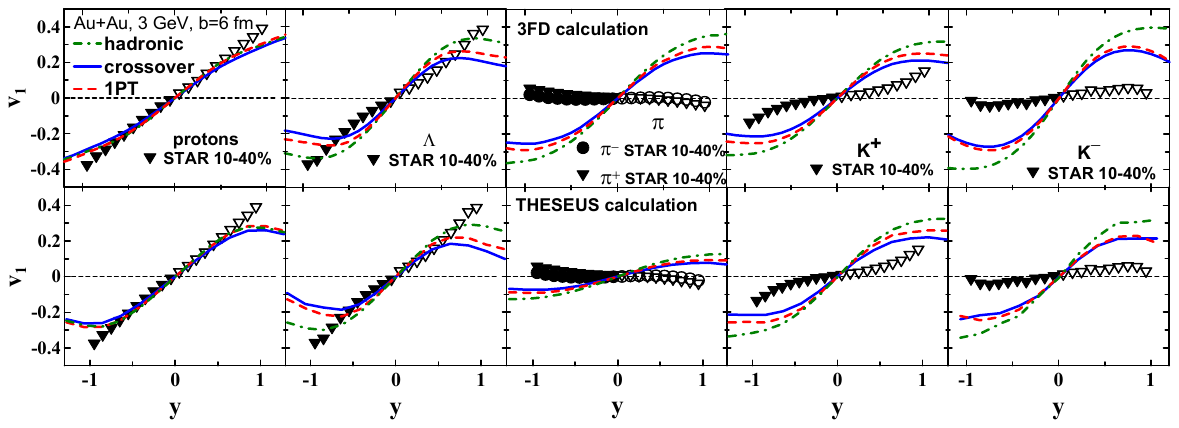}
  \caption{%(Color online)
Directed flow of  
protons, pions, $\Lambda$ hyperons, and kaons 
as function of rapidity in semicentral ($b=$ 6 fm)    
Au+Au collisions at collision energy of $\sqrt{s_{NN}}=$ 3 GeV. 
Results are calculated within the 3FD model (upper row of panels) and 
the THESEUS (lower row of panels) with hadronic, 1PT, and crossover EoSs. 
STAR data are from Ref. \cite{STAR:2021yiu}.
}
    \label{fig:v1-STAR-fxt-3GeV-3FD}
\end{figure*}
\begin{figure*}[!tbh]
\includegraphics[width=.91\textwidth]{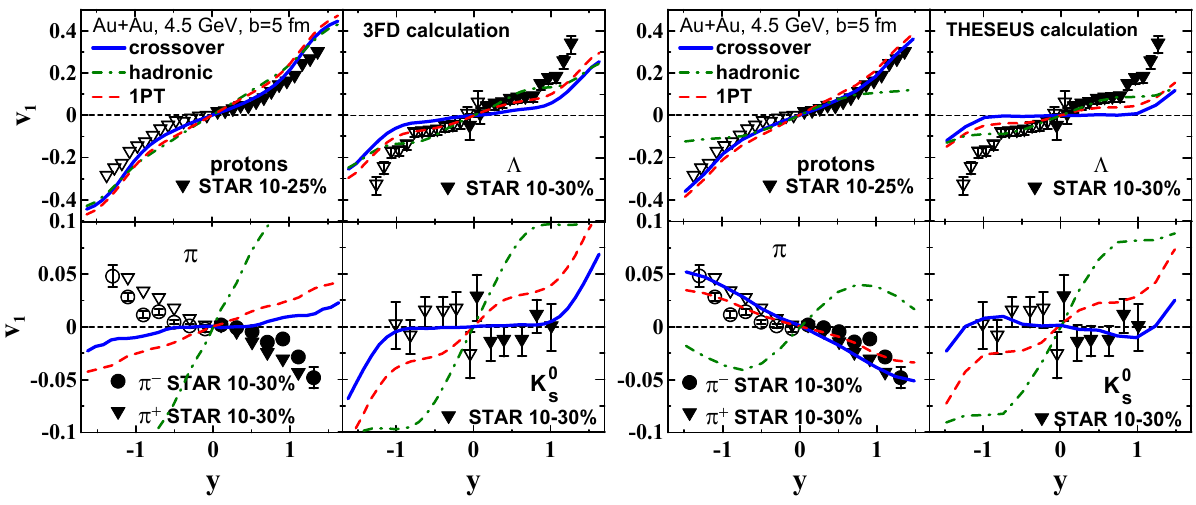}
  \caption{%(Color online)
Directed flow of  
protons, pions, $\Lambda$ hyperons, and kaons ($K^0$ short)
as function of rapidity in semicentral ($b=$ 5 fm)    
Au+Au collisions at collision energy of $\sqrt{s_{NN}}=$ 4.5 GeV. 
Results are calculated within the 3FD model (left block of panels) and 
the THESEUS (right block of panels) with hadronic, 1PT, and crossover EoSs. 
STAR data are from Ref. \cite{STAR:2020dav}.
}
    \label{fig:v1-STAR-fxt-45GeV-3FD}
\end{figure*}

The calculated directed flow of  
protons, pions, $\Lambda$ hyperons, and (anti)kaons 
as function of rapidity in semicentral     
Au+Au collisions at collision energies of $\sqrt{s_{NN}}=$ 3 and 4.5 GeV   
are presented in 
Figs. \ref{fig:v1-STAR-fxt-3GeV-3FD} and \ref{fig:v1-STAR-fxt-45GeV-3FD}, respectively.
These calculations were performed in the 3FD model without any afterburner and 
within THESEUS (i.e. with the UrQMD afterburner).  
The collision centrality was associated with the corresponding mean impact parameter  by means of 
the Glauber simulations based on the nuclear overlap calculator \cite{web-docs.gsi.de}. 
Of course, this an approximate way to simulate the experimental centrality selection. 
Nevertheless, it captures the main trends of directed flow.
The results are compared with STAR data \cite{STAR:2021yiu,STAR:2020dav}.

As seen, the proton $v_1$ flow is well reproduced with and without afterburner. The afterburner slightly
improves the description at 4.5 GeV, while worsens it at forward/backward rapidities at 3 GeV without 
changing the midrapidity slope. 
The midrapidity proton flow turns out to be almost independent of the used EoS even at 4.5 GeV, 
where the QGP transition already takes place, see Fig. \ref{fig:TnB-box_2024}.  
This is because the proton flow is formed at the 
early stage of the collision \cite{Baumgardt:1975qv,Stoecker:2004qu,Poskanzer:1998yz}. 
At considered collision energies, this stage is developed in the hadronic phase for all considered EoSs, 
see Fig. \ref{fig:TnB-box_2024}, where all considered EoSs are very similar, see Fig. \ref{fig:3-EOS-2024}.
Moreover, the stopping power of the matter, i.e. friction forces of the 3FD model \cite{Ivanov:2013wha}, 
are identical in the hadronic phase for all considered scenarios. Consequently, the flow appears to be 
quite independent of the used EoS even at 4.5 GeV. Note that the proton directed flow does depend on the 
EoS at the BES RHIC energies \cite{Konchakovski:2014gda,Ivanov:2014ioa,Ivanov:2016sqy}, 
where the transition to QGP occurs already the early stage of the collision.

\begin{figure}[!tbh]
\includegraphics[width=.46\textwidth]{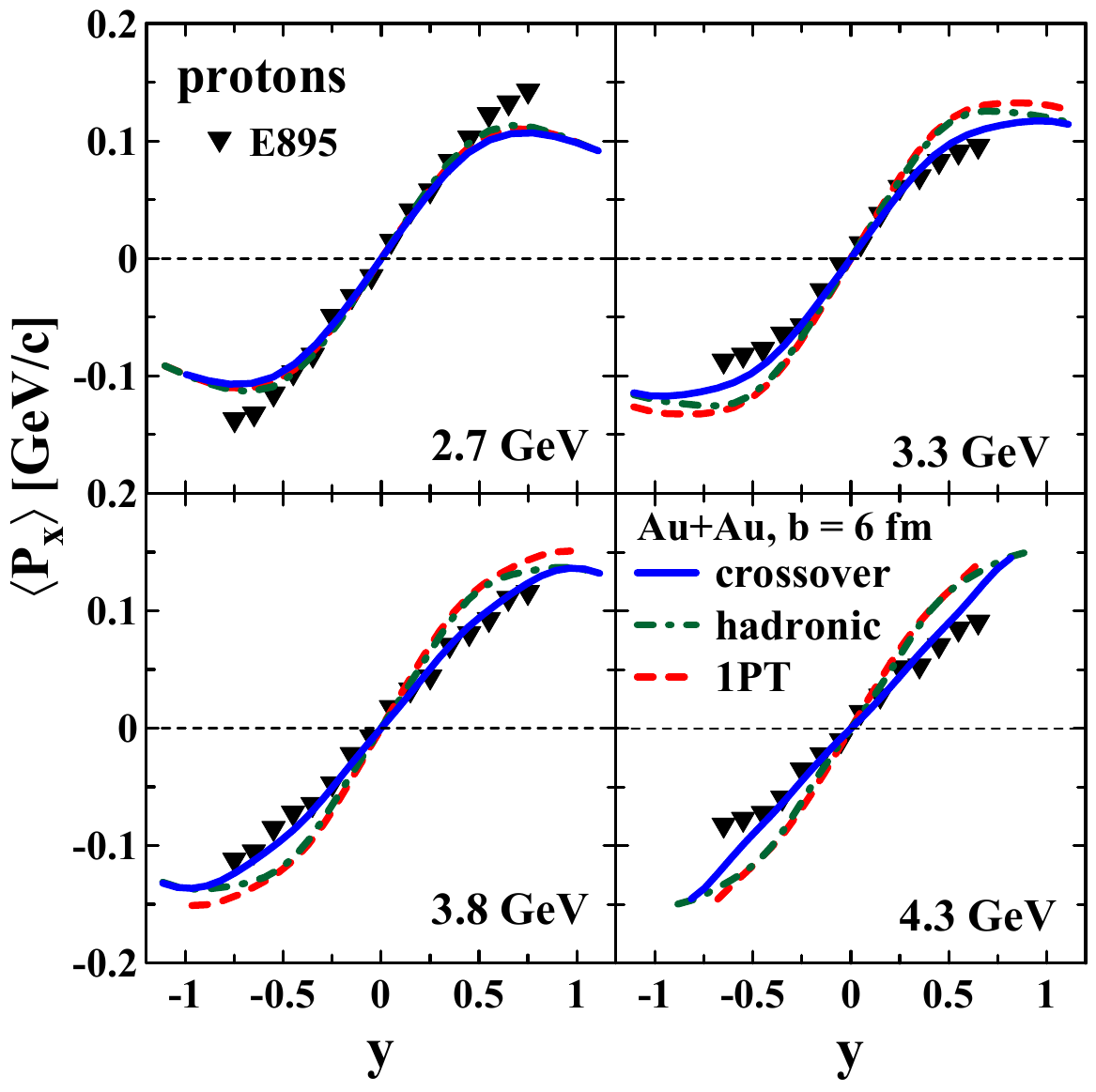}
  \caption{%(Color online)
Transverse flow of protons
as function of rapidity in semicentral ($b=$ 6 fm)    
Au+Au collisions at collision energies of $\sqrt{s_{NN}}=$ 2.7--4.3 GeV
($E_{\rm lab}=$ 2$A$, 4$A$, 6$A$, and 8$A$ GeV). 
Results are calculated within the 3FD model with hadronic, 1PT, and crossover EoSs. 
E895 data are from Ref. \cite{E895:2000maf}.
}
    \label{fig:Au2-8gas_Px-gt}
\end{figure}
Figure \ref{fig:Au2-8gas_Px-gt}
illustrates the description of the old E895 data \cite{E895:2000maf} in the same collision energy range. 
These data are presented in terms of transverse flow defined 
as \cite{Danielewicz:1985hn}
\begin{eqnarray}
\langle P_x\rangle (y)= 
\frac{\displaystyle \int d^2 p_T \ p_x \ E \ dN/d^3p }%
{\displaystyle \int d^2 p_T\ E \ dN/d^3p}, 
\label{eq-px}
\end{eqnarray}
where $p_x$ is the transverse momentum of the proton in the reaction
plane, $E \ dN/d^3p$ is the invariant momentum distribution of protons
with $E$ being the proton  energy, 
and integration runs over the transverse momentum $p_T$. 
This is done because the E895 data in terms of $v_1 (y)$ raised many questions, 
as it was discussed in Ref. \cite{Ivanov:2014ioa} in detail. 
In addition, they contradict the new STAR-FXT data. 
As seen from Fig. \ref{fig:Au2-8gas_Px-gt}, the crossover EoS gives almost perfect description 
of the proton transverse flow in the midrapidity regions. 
This gives hope that the future STAR-FXT proton data at the energies between 3 and 4.5 GeV
will be also well reproduced with the crossover scenario. 
The dependence on the EoS is quite moderate, similar to that at 3 and 4.5 GeV in 
Figs. \ref{fig:v1-STAR-fxt-3GeV-3FD} and \ref{fig:v1-STAR-fxt-45GeV-3FD}.

The $\Lambda$ flow turns out to be more sensitive to the EoS,  
see Figs. \ref{fig:v1-STAR-fxt-3GeV-3FD} and \ref{fig:v1-STAR-fxt-45GeV-3FD}, 
because $\Lambda$s are produced 
in highly excited but still baryon-rich regions of the colliding system. 
Note that the same freeze-out energy density can be achieved by means of either 
high baryon density at moderate temperature or high temperature at moderate baryon density, 
the latter we refer as the highly excited but still baryon-rich regions. 
These regions are formed later, 
when the temperature reaches high values, see Fig. \ref{fig:TnB-box_2024}. 
The afterburner stronger affects evolution in these regions. 
It reduces the midrapidity slope of the $\Lambda$ flow, 
making the crossover EoS somewhat preferable at 3 GeV, 
while the hadronic EoS turns out to be preferable at 4.5 GeV. 
In view of this sensitivity to the afterburner, 
definite conclusions on the EoS relevance can hardly be made based on the $\Lambda$ flow.

The meson flow probes dynamics in highly excited baryon-rich and baryon-depleted regions of the system. 
Again, the same freeze-out energy density can be achieved by means of either high baryon density at moderate temperature, or high temperature at moderate baryon density, as is the case in the baryon-depleted regions.
The highly excited baryon-depleted regions are formed even later than the excited baryon-rich regions, 
when the transverse expansion already dominates. 
Relative contributions of the baryon-rich and baryon-depleted regions to nucleon and meson production 
may be different. 
Therefore, the mesonic flow does not necessary follow the nucleon pattern. 
Nevertheless, the mesonic flow is very similar to the baryon one after the 3FD stage at 3 GeV, see 
Fig. \ref{fig:v1-STAR-fxt-3GeV-3FD}, 
which indicates that mesons are mostly produced from decays of baryonic resonances.

At 4.5 GeV, the mesonic flow substantially differ from the baryon one, see
Fig. \ref{fig:v1-STAR-fxt-45GeV-3FD}. 
This difference at 4.5 GeV concerns only the EoSs involving the transition to the QGP, 
whereas the hadronic EoS results in the mesonic flow being similar to the baryon one. 
This implies that in the QGP-transition scenarios the relative contribution of 
the baryon-depleted regions becomes higher because of thermal production of mesons and mesonic resonances. 
The EoS becomes softer in the QGP and hence the pressure causing the directed flow is reduced. 
To a greater extent, this concerns the mesonic flow formed at the QGP stage of the collision.
Therefore, the mesonic flow may indicate the transition to the QGP. 
However, there are other circumstances that may prevent us from drawing definite conclusions from the mesonic flow.
One of them is the afterburner.

The pion flow is strongly affected by the afterburner. If the 3FD-calculated pion flow hardly resembles the 
corresponding data, after the afterburner stage, it almost perfectly describes these data 
at 4.5 GeV within the crossover and 1PT scenarios. 
The hadronic scenario evidently fails to reproduce the pion flow at 4.5 GeV. 
The afterburner even changes the sign of the midrapidity slope of the crossover and 1PT flows at 4.5 GeV. 
Note that the 3FD model and hence THESEUS do not distinguish positive, neutral, and negative pions. 
Therefore, the calculated pion flow refers to the flow of all pions. This strong dependence on the afterburner
is a consequence of the shadowing discussed in Sec. \ref{3FD and THESEUS}. 
At 3 GeV, the afterburner shifts the 3FD-calculated flow closer to the data 
but still not enough to reproduce them. This insufficient effect 
of the afterburner is a result of the shortcoming of the THESEUS afterburner 
discussed in Sec. \ref{3FD and THESEUS}: The afterburner skips the stage of shadowing
the afterburner expansion of the central fireball by spectators still being in the hydrodynamic phase.
At 4.5 GeV, this skipped stage is already of miner importance because 
the time for the nuclei to pass each other 
becomes shorter relative to the time scale of the participant evolution.

The $K^+$ directed flow is not changed by the afterburner stage because of small cross sections 
of their interactions with other hadrons. Indeed, the kaon-nucleon cross section is about 10 mb, 
while the nucleon-nucleon one is about 40 mb \cite{Hartnack:2011cn,Bleicher:1999xi,Song:2020clw}. 
Indeed, at somewhat lower energies (2$A$ GeV) it was concluded that 
after their production the K-mesons suffer not more than one rescattering before escaping
\cite{Zwermann:1984pq,Russkikh:1992tg}. 
At the energy of 3 GeV ($\approx$ 3$A$ GeV), higher densities are achieved in the collisions and 
therefore rescatterings become more frequent, however not frequent enough to thermalize the kaon.
This explains why the calculated flow 
(on the assumption of the kaon thermalization) is so different from the data, see Fig. 
\ref{fig:v1-STAR-fxt-3GeV-3FD}.

At slightly higher collision energy of 3.85 GeV (6$A$ GeV), the rescatterings of the kaons with
the nucleons in the dense matter already cause them to flow
in the direction of the nucleons, as it was reported in Ref. \cite{Pal:2000yc}. 
Thus, the transition from rare-collisional to collisional regime occurs in the considered energy range. 
At the energy of 4.5 GeV, the kaons can be considered well thermalized at the late stage of nuclear collision, 
however the afterburner still does not affect the flow, as seen from Fig.   
\ref{fig:v1-STAR-fxt-45GeV-3FD}.

At the same time, the $K^-$ directed flow at 3 GeV
is reduced by the afterburner because the $NK^-$ cross section is of the order of 
40 mb or even higher at low relative $NK^-$ energies \cite{Hartnack:2011cn,Bleicher:1999xi,Song:2020clw}. 
However, the calculated $K^-$ flow is still essentially stronger than the experimental one. 
Apparently, this is a result of the aforementioned shortcoming of the THESEUS afterburner, i.e.  
a lack of shadowing of the central fireball by still hydrodynamically evolving spectators.

\begin{figure*}[!htb]
\includegraphics[width=.7\textwidth]{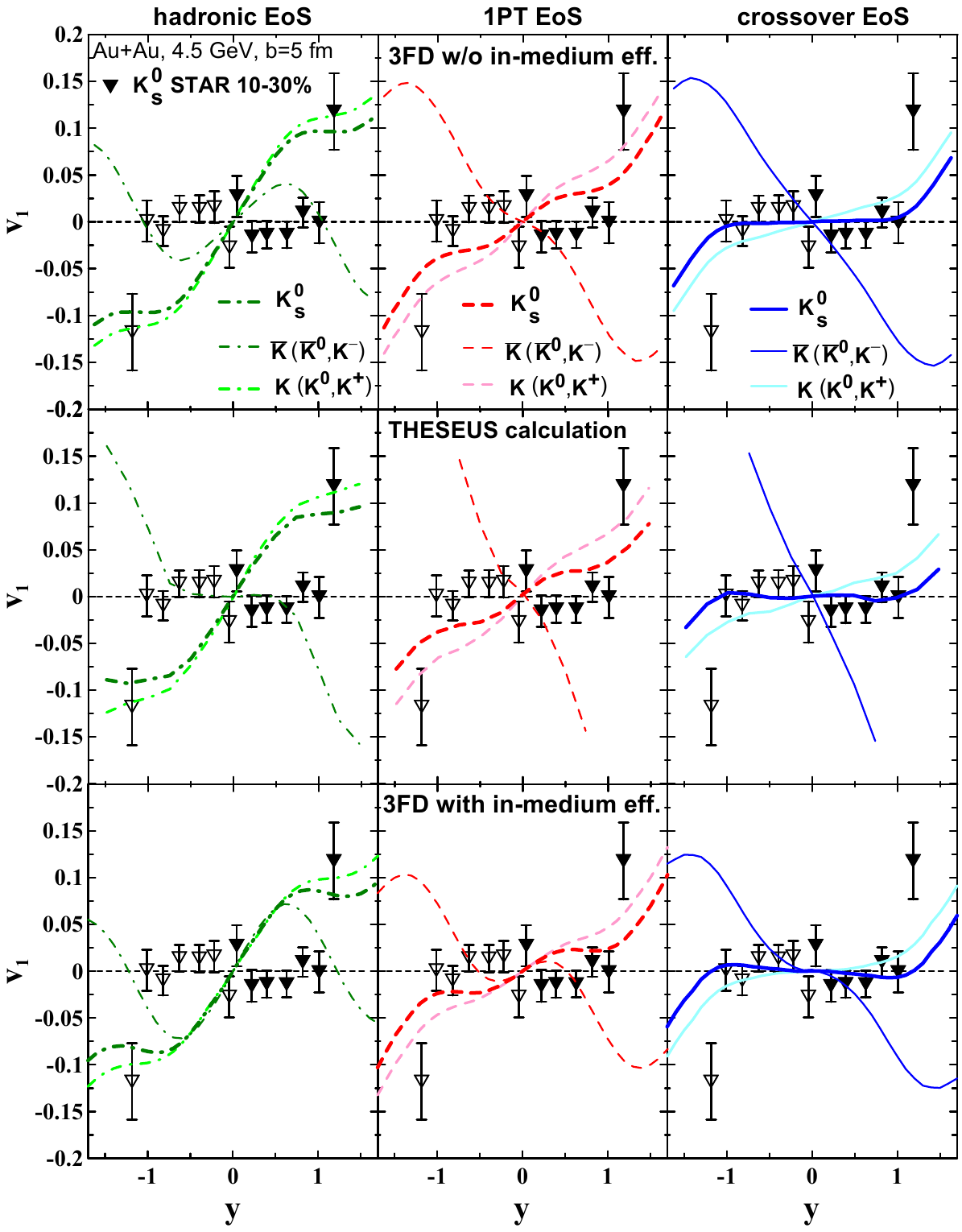}
  \caption{%(Color online)
Directed flow of kaons, antikaons and  $K^0_s$ mesons 
as function of rapidity in semicentral ($b=$ 5 fm)    
Au+Au collisions at collision energy of $\sqrt{s_{NN}}=$ 4.5 GeV. 
Results are calculated within THESEUS (the middle row of panels) and also within the 3FD model 
with (the lower row of panels) and without (the upper row of panels) in-medium modifications of (anti)kaons. 
%with hadronic, 1PT, and crossover EoSs 
STAR data are from Ref. \cite{STAR:2020dav}.
}
    \label{fig:v1-STAR-fxt-45GeV-5fm-Ks+Kseff}
\end{figure*}

Within the 3FD model, 
we calculate $v_1 (y)$ for $K^0_s$ mesons in terms of those for $K^0$ and $\bar{K}^0$ as follows
\begin{eqnarray}
\label{K0s}
v_1^{K0s}(y) &=& \left(v_1^{K0}(y)\frac{dN^{K0}}{dy} + v_1^{\overline{K}0}(y)\frac{dN^{\overline{K}0}}{dy}\right)
\cr
&/&
\left(\frac{dN^{K0}}{dy} + \frac{dN^{\overline{K}0}}{dy}\right), 
\end{eqnarray}
where $dN^{K0}/dy$ and $dN^{\overline{K}0}/dy$ are rapidity distributions of the $K^0$ and $\bar{K}^0$ mesons.
Eq. (\ref{K0s}) does not imply that $K^0_s$ consists of $K^0$ and $\bar{K}^0$ in this proportion. It only means 
that $K^0_s$ mesons originate from $K^0$ and $\bar{K}^0$ mesons that are emitted from the interaction region. 
These $K^0$ and $\bar{K}^0$ mesons keep their momenta and thus their flow pattern
after escaping from the interaction region. Therefore, the corresponding fractions of produced $K^0_s$ mesons
carry these $K^0$ and $\bar{K}^0$ flow patterns. 
The $\bar{K}^0$ number is about 20\% of that of $K^0$ at 4.5 GeV.

The directed flow of $K^0_s$ mesons at  4.5 GeV strongly depends on the EoS and moderately depends on 
the afterburner. This moderate dependence on the afterburner is a consequence of the large fraction of 
$K^0$ mesons in produced $K^0_s$. The $K^0$ mesons are practically unaffected by the afterburner. 
Therefore, the $K^0_s$ directed flow is a good probe of the hot and dense stage of the collision. 
As seen from Fig. \ref{fig:v1-STAR-fxt-45GeV-3FD}, the crossover EoS is certainly preferable 
for reproduction of the data.

Thus, the directed flows of various particles provide information on dynamics in various parts 
and at various stages of the colliding system depending on the particle. 
However, the information on the EoS is not always directly accessible 
because of strong influence of the afterburner stage or insufficient thermalization of kaons.  
The crossover scenario gives the best overall description of the data, of course, 
with all reservations regarding the above-mentioned difficulties in applying the model.

\section{Directed Flow of Kaons} 
  \label{Directed Flow of Kaons}

The kaons deserve a separate discussion. As has been mentioned above, 
the afterburner does not affect the flow of kaons because of small cross sections 
of their interactions with other hadrons but noticeably changes  the antikaon flow at 3 GeV,   
see Fig. \ref{fig:v1-STAR-fxt-3GeV-3FD}. It is instructive to consider the kaon and antikaon flows 
at 4.5 GeV, in spite of absence of the corresponding data. 

The directed flow of kaons, antikaons and  $K^0_s$ mesons 
as function of rapidity in semicentral ($b=$ 5 fm)    
Au+Au collisions at collision energy of $\sqrt{s_{NN}}=$ 4.5 GeV is presented 
in Fig. \ref{fig:v1-STAR-fxt-45GeV-5fm-Ks+Kseff}. 
The flows of kaons and antikaons are marked as $(K^0,K^+)$ and $(\bar{K}^0,K^-)$, respectively, 
because the 3FD model does not distinguish the corresponding mesons. 
The kaon flow again turns out to be insensitive to the afterburner.

The flow of antikaons
is enhanced by the afterburner, contrary to the reduction of the antikaon flow at 3 GeV. 
Notably, midrapidity slopes of the kaon and antikaon flow are of the opposite sign 
for the crossover and 1PT EoSs while they are both non-negative within the hadronic scenario. 
Apparently, the antiflow of the antikaons is again related to the aforementioned shadowing of 
the decay of central blob by the spectator matter. This shadowing is present already in the 
3FD stage of the evolution, as seen from the upper row of panels in  
Fig. \ref{fig:v1-STAR-fxt-45GeV-5fm-Ks+Kseff}. The afterburner additionally enhances this 
shadowing and hence the antiflow, 
see the middle row of panels in  Fig. \ref{fig:v1-STAR-fxt-45GeV-5fm-Ks+Kseff}. 
These opposite signs of the  midrapidity slopes of the kaon and antikaon flows
can be considered as a prediction for the flow at 4.5 GeV.

In-medium modifications of kaons are discussed in connection with  
chiral symmetry restoration and neutron star properties, see review \cite{Tolos:2020aln}.  
In Refs. \cite{Pal:2000yc,Hartnack:2011cn,Song:2020clw,Cassing:2014xca}, 
it was found that in-medium modifications of kaons are very important for description of the  
kaon observables, in particular, the kaon directed flow. 
It was reported that these in-medium effects can even change the midrapidity slope of the 
kaon flow at the energy of 3.85 GeV (6$A$ GeV) \cite{Pal:2000yc}, i.e. in the energy range we consider here.

In the relativistic mean-field approximation for the baryon degrees of freedom \cite{Li:1994cu,Schaffner:1994bx}, 
the in-medium (anti)kaon energy reads
\begin{eqnarray}
\label{omek}
E({\bf p})=\left[m_K^2+{\bf p}^2-
\frac{\Sigma_{KN}}{f^2_K} \rho +\left(\frac{3}{8}\frac{n}{f^2_K}\right)^2
\right]^{1/2} \!\!\!\!
\pm \frac{3}{8}\frac{n}{f^2_K},
\end{eqnarray}
where ${\bf p}$ is the three-momentum of the (anti)kaon, 
the upper(lower) sign refers to $K$($\bar{K}$). 
\begin{eqnarray*}
\label{rho}
n &=& \sum_B\langle{\bar B}\gamma^0 B\rangle, 
\cr 
\rho &=& \sum_B\langle{\bar B}B\rangle
\end{eqnarray*}
are the proper baryon density and scalar baryon density, respectively, which are sums over 
various baryons $B$.  
Numerical values of the kaon decay constant, $f_K=106$ MeV, and 
the kaon-nucleon sigma term, $\Sigma_{KN} = 350$ MeV, are taken from Ref. \cite{Hartnack:2011cn}. 
The term proportional to $\Sigma_{KN}$ results from the attractive scalar interaction
due to explicit chiral symmetry breaking.

The above expression was derived for the so-called $s$-wave interaction. 
Importance of $p$-wave kaon-baryon interactions was indicated in Refs. \cite{Kolomeitsev:1995xz,Kolomeitsev:2002pg}. 
The treatment of the kaon-baryon interaction beyond the mean-field approximation, i.e. with the 
G-matrix approach  \cite{Song:2020clw,Tolos:2020aln}, also turned out to be important. 
Therefore,  Eq. (\ref{omek}) can only serve as a basis for the estimation of the in-medium effects in (anti)kaon production. 
Below we use Eq. (\ref{omek}) for this purpose. 
The same form of the in-medium kaon energy was used in Ref. \cite{Pal:2000yc}. 

The version of the UrQMD that is implemented in THESEUS is not suitable for treatment of the medium modified kaons. 
Therefore, we study the in-medium effects within the 3FD model. 
The masses and chemical potentials of (anti)kaons were modified at the freeze-out stage. 
%Apparently, this gives a lower estimate of the in-medium effects for kaons because they are frozen out earlier 
%then other hadrons and therefore the in-medium effect is accumulated for a longer time, i.e. throughout the evolution of the system. 
Results of the 3FD calculation of the directed flow taking into account the in-medium kaon modification 
at collision energy of $\sqrt{s_{NN}}=$ 4.5 GeV are shown  
in the lower row of panels of Fig. \ref{fig:v1-STAR-fxt-45GeV-5fm-Ks+Kseff}. 
As seen, the effect of this in-medium modification is quite moderate. However, it slightly improves the 
agreement with the $K^0_s$ data within the 1PT EoS and especially the crossover scenario. 
This improvement is practically the same as that resulted from the afterburner. 
It is remarkable that 
the change of the antikaon flow due to the in-medium effects is opposite to that caused by the afterburner.

\begin{figure}[!htb]
\includegraphics[width=.47\textwidth]{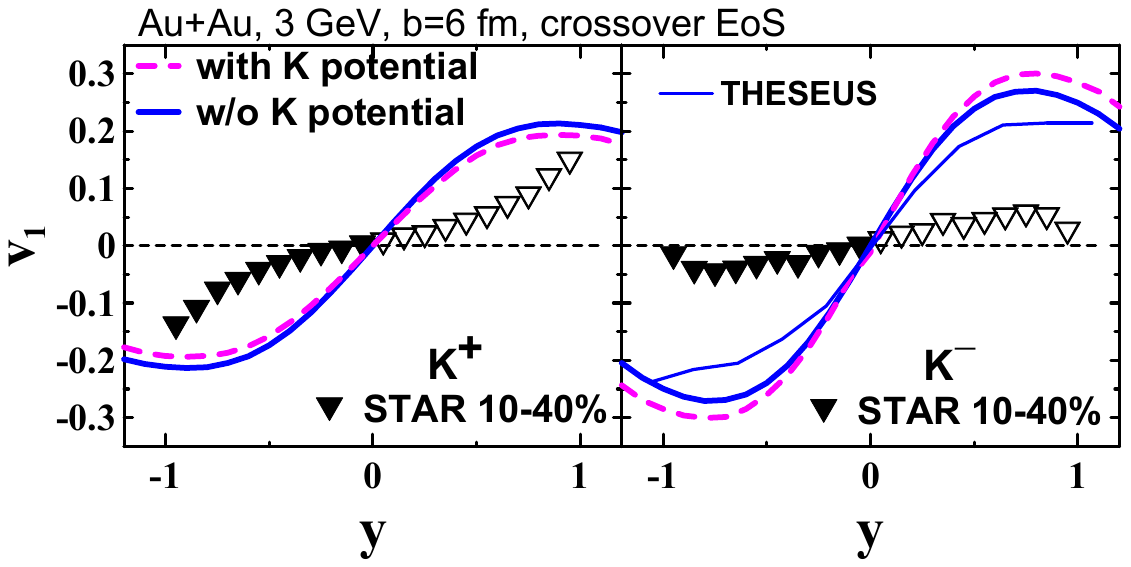}
  \caption{%(Color online)
Directed flow of kaons (left panel) and antikaons (right panel) 
as function of rapidity in semicentral ($b=$ 6 fm)    
Au+Au collisions at collision energy of $\sqrt{s_{NN}}=$ 3 GeV. 
Results are calculated  with (with $K$ potential) and without (w/o $K$ potential) 
in-medium modifications of (anti)kaons  
for the crossover EoSs.  
STAR data are from Ref. \cite{STAR:2020dav}.
}
    \label{fig:v1-STAR-fxt-Keff-3GeV}
\end{figure}

As seen from Fig. \ref{fig:v1-STAR-fxt-Keff-3GeV}, the effect of the in-medium modifications of (anti)kaons 
is also small contrary to that found in Refs. \cite{Pal:2000yc,Hartnack:2011cn,Song:2020clw,Cassing:2014xca}. 
Apparently, this is because the kaons were incompletely equilibrated in the matter in kinetic simulations of 
Refs. \cite{Pal:2000yc,Hartnack:2011cn,Song:2020clw,Cassing:2014xca} and hence the in-medium effect 
was accumulated throughout the evolution of the system. In the 3FD simulations they are 
completely equilibrated and the in-medium modifications appear only at the freeze-out,  
leaving them insufficient time to manifest themselves. 
Therefore, the present calculation should be considered as a lower estimate of the in-medium effects for kaons.  
Again the in-medium modifications and the afterburner result in opposite changes in 
the antikaon directed flow. Only the afterburner decreases and the in-medium modification increases the flow, 
contrary to that at 4.5 GeV. This is because there is the normal flow at 3 GeV instead of antiflow at 4.5 GeV.

We can conclude that the directed flow of kaons or $K^0_s$ is a promising probe of the EoS at hot and dense stage 
of the collision at 4.5 GeV because it is not affected by the afterburner stage.  
At 3 GeV, the kaons do not appear to be fully equilibrated in matter and therefore 
do not reflect the EoS of the matter. 
The antikaon flow is also a good EoS probe, which is however strongly modified during the afterburner evolution. 

%_________________________________________________________________
\section{Summary}
\label{Summary}

The directed flow of various hadrons 
at energies $\sqrt{s_{NN}}=$ 3 and 4.5 GeV were calculated and compared  with recent STAR-FXT data 
\cite{STAR:2020dav,STAR:2021yiu}. 
The calculations were performed within the 3FD model 
\cite{Ivanov:2005yw,Ivanov:2013wha} and also within the
THESEUS generator \cite{Batyuk:2016qmb,Batyuk:2017sku,Kozhevnikova:2020bdb} in order to  
to study the effect of the UrQMD afterburner stage on the directed flow. 
Three different EoSs are used in the 
simulations: a purely hadronic EoS \cite{gasEOS} and two EoSs
with deconfinement transitions \cite{Toneev06}, i.e. 
an EoS with a strong first-order phase transition and one with a
smooth crossover transition.

At these collision energies, the time for the nuclei to pass each other
is long relative to the time scale of the participant evolution and therefore 
the interaction between participants and spectators (shadowing) is important.
In particular, the squeeze-out effect is a consequence of this shadowing. 
This shadowing only partially is taken into account within the 3FD evolution because the central
fireball remains to be shadowed even after the freeze-out. Therefore, the afterburner stage 
becomes of prime importance.

The afterburner shifts the 3FD-calculated flow closer to the data 
but still not enough to reproduce the pion and antikaon flow at 3 GeV. This insufficient effect 
of the afterburner results from shortcoming of its isochronous initialization: 
The afterburner skips the stage of shadowing of
the post-freeze-out expansion of the central fireball by spectators still hydrodynamically evolving.

The directed flows of various particles provide information on dynamics in various parts 
and at various stages of the colliding system depending on the particle. 
However, the information on the EoS is not always directly accessible 
because of strong influence of the afterburner stage or insufficient equilibration,  
as it happens with kaons at 3 GeV. 
Based on these simulations, the following conclusions were drawn:

\begin{itemize}

	\item 

The proton flow is formed at the early stage of the collision, where the matter is not yet equilibrated. 
Therefore, it probes the properties of this nonequilibrium  matter rather than its EoS that 
implies the equilibrated matter. The proton flow is well reproduced within all three considered scenarios 
and is practically independent of the afterburner.

	\item 

The $\Lambda$ flow turns out to be more sensitive to the
EoS because $\Lambda$s are produced in highly excited but still
baryon-rich regions of the colliding system. These regions
are formed later, when the temperature reaches high values. 
The afterburner stronger affects evolution in these regions.

	\item 

The meson flow probes dynamics of highly excited baryon-rich and baryon-depleted regions of the system.
The highly excited baryon-depleted regions are formed even later than the excited baryon-rich regions, 
when the transverse expansion already dominates.

	\item 

The pion flow is strongly affected by the afterburner.
This strong dependence on the afterburner is a consequence of
the shadowing. 

	\item 

The directed flow of kaons or $K^0_s$ is a promising probe of the EoS at hot and dense stage 
of the collision at 4.5 GeV because it is not affected by the afterburner stage.  
At 3 GeV, the kaons do not appear to be fully equilibrated in matter and 
therefore do not reflect the EoS of the matter. 
The antikaon flow is also a good EoS probe, which is however strongly modified during the afterburner evolution.

\end{itemize}

In conclusion, the crossover scenario gives the best overall description of the data, of course, 
with all reservations regarding the above-mentioned difficulties in applying the model. 
This crossover EoS is soft in the hadronic phase. 
This result agrees with that in Refs. \cite{Ivanov:2014ioa,Ivanov:2016sqy,Pal:2000yc,Danielewicz:2002pu}
but is in contrast to Refs. 
\cite{Nara:2020ztb,Oliinychenko:2022uvy,Steinheimer:2022gqb,OmanaKuttan:2022aml,Wu:2023rui,Yao:2023yda}, 
where stiff  EoSs were found being preferable for the reproduction of the directed flow at 3 GeV. 
The conclusion about the preference of the stiff EoS  
\cite{Nara:2020ztb,Oliinychenko:2022uvy,Steinheimer:2022gqb,OmanaKuttan:2022aml,Wu:2023rui,Yao:2023yda}, 
was mostly based on the proton flow, which is formed at the
early nonequilibrium stage of the collision.  
Therefore, the proton flow is a combined result of the EoS and the stopping power of the matter. 
Different combinations of the EoS and the stopping power can properly describe the proton flow. 
Directed flows of different hadrons, as well as other bulk observables 
should be considered together to decouple effects of the EoS and the stopping power. 

Within the preferred crossover scenario, 
the transition into QGP in Au+Au collisions occurs at collision energies between 3 and 4.5 GeV,  
at baryon densities $n_B \gsim 4 n_0$ and temperatures $\approx 150$ MeV.  
This implies that the EoS additionally softens at 4.5 GeV. 
This softening and hence transition into QGP at 4.5 GeV well agrees with conclusions made 
in Refs. 
\cite{Nara:2020ztb,Oliinychenko:2022uvy,Steinheimer:2022gqb,OmanaKuttan:2022aml,Wu:2023rui,Yao:2023yda}.

% ________________________________________________________________
\begin{acknowledgments}

Fruitful discussions with  D.N. Voskresensky are gratefully acknowledged.
This work was carried out using computing resources of the federal collective usage center ``Complex for simulation and data processing for mega-science facilities'' at NRC "Kurchatov Institute" \cite{ckp.nrcki.ru}.
and computing resources of the supercomputer "Govorun" at JINR \cite{govorun}. 

\end{acknowledgments}
% ________________________________________________________________

\input{v1-STAR-fxt-bib-rev-hep.tex}

%%%%%%%%%%%%%%%%%%%%%%%%%%%%%%%%%%%%%%%%%%%%%%%%%%%%%
\end{document}

%% file: v1-STAR-fxt-bib-rev-hep.tex
% ________________________________________________________________